\documentclass[12pt, a4paper]{article}
\usepackage[font=footnotesize]{caption}
\usepackage[utf8]{inputenc}
\usepackage{graphicx}
\usepackage{authblk}
\usepackage{booktabs}
\usepackage{verbatim}
\usepackage{url}
\usepackage{caption}
\usepackage{subcaption}
\usepackage{indentfirst}
\usepackage{url}
\usepackage{amssymb}
\usepackage{amsmath}

\usepackage{multirow}
\usepackage[table,xcdraw]{xcolor}

\usepackage{todonotes}

\usepackage{rotating}
\usepackage{array}
\title{Detecting Hierarchical Clusters and \\ Estimating their Modularity \\ Directly from Dendrograms}
\author{Alexandre Benatti$^1$ and \\ Luciano da F. Costa$^2$}

\affil{
$^1$Institute of Mathematics and Statistics - DCC \\
University of S\~ao Paulo \\
Rua do Mat\~ao, 1010, \\ S\~ao Paulo, SP 05508-090 Brazil 
\\ \vspace{0.5cm}
$^2$S\~ao Carlos Institute of Physics - DFCM \\
University of S\~ao Paulo \\
Av. Trabalhador S\~ao-Carlense, 400, \\ S\~ao Carlos, SP 13566-590 Brazil \\
(Prof. Senior)
}

\date{\emph{22nd May, 2026}}

\begin{document}

\maketitle

\begin{abstract}
Identifying possible clusters in datasets and estimating their hierarchical modularity are central tasks in pattern recognition. In the present work, concepts and methodologies are described for performing these tasks while considering only the density of mergings obtained from hierarchical representations (dendrograms) of data inter-relationship along a scale variable. More specifically, the mergings of subclusters along the scale variable are obtained, yielding a respective merging density function. After this function is equalized along the scale variable, peak detection is applied in order to estimate, within a specified resolution, the main hierarchical levels and their clusters. After quantifying infinitesimal modularity of the dendrogram at a fixed scale value, taking into account the uniformity of the size of the identified clusters and their average size, the overall, average, and group hierarchical modularities are obtained. The potential of the reported approach is illustrated for some types of data and dendrograms and additive dendrograms obtained from DLA patterns, and the possibility of recursive cluster detection is also considered.
\end{abstract}

\section{Introduction}\label{sec:introduction}

Finding groups of elements in data sets characterized by respective features constitutes one of the main tasks in pattern recognition and machine learning and its various applications. This task can be performed supervised or unsupervised (e.g.~\cite{duda2000pattern, theodoridis2006pattern}), the latter type being characterized by a lack of preliminary information about the possible presence of groups and about their properties.

Among the approaches that have been described for unsupervised pattern recognition, or \emph{clustering} (e.g.~\cite{hall1991,jain1988algorithms,jain1999data,rodriguez2019}), hierarchical methods have allowed the data elements to be inter-related along a dissimilarity (or distance) scale parameter, henceforth referred to as $s$, which can express the distance or dissimilarity between the data features of pairs of data elements. Estimation of the hierarchical relationship between these elements can be obtained by starting with the individual data elements as leaves and progressing through mergings into subgroups until a single cluster is obtained. The so-obtained hierarchical relationships are often represented in terms of \emph{dendrograms} (e.g.~\cite{hill1980,duda2000pattern,theodoridis2006pattern,de2013pattern}), which provide a synthetic and effective indication of the relationship between the original data elements for a range of interconnectivity scales.

It is interesting to observe that, strictly speaking, trees are topological structures characterized by a hierarchy of branches, not necessarily taking into account geometrical aspects including distances or similarities. In the present work, trees incorporating metrics are called \emph{dendrograms}. Therefore, dendrograms can be understood as a special type of trees.

For simplicity's sake, the topology of the dendrograms considered in this work are assumed to correspond to a binary tree, but the approach can be simply adapted to more general situations. It should also be observed that because dendrograms typically do not preserve all the information available in the original data features, they do not establish an invertible mapping with the original data, so that the data features cannot be recovered from a dendrogram. Therefore, a same dendrogram can be obtained from two or more distinct data sets.

\begin{figure}[h]
  \centering
     \includegraphics[width=0.8 \textwidth]{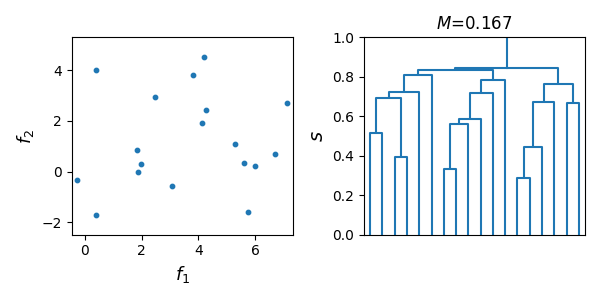}\\\vspace{-.1 cm}
     \hspace{1.1 cm} (a) \hspace{4.3 cm} (b)
     \includegraphics[width=0.8 \textwidth]{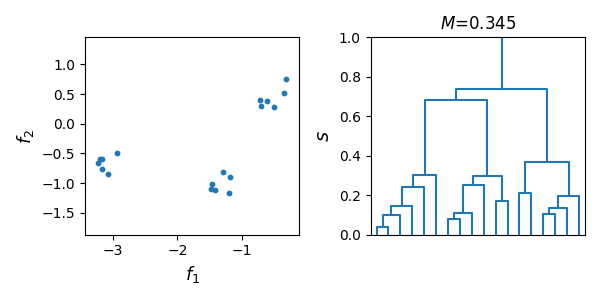}\\\vspace{-.1 cm}
     \hspace{1.1 cm} (c) \hspace{4.3 cm} (d)
 \caption{Example of a less modular (a) and a more modular (c) data sets, with the x-y coordinates taken as features of the data elements, together with their respective dendrograms (b, d) obtained by average linkage criterion. The modularity of the dendrogram in (d) involves two main scales of inter-relationship corresponding to the distances among the elements of each cluster as well as the distances among the clusters themselves. Other dendrograms may present additional main hierarchical levels.}\label{fig:modularity}
\end{figure}

Even though dendrograms do not directly incorporate the original data features, they still provide comprehensive information about the hierarchical relationship between the original data elements to the point that it becomes interesting to try to estimate the presence of possible clusters and their modularity directly from dendrograms. This possibility is illustrated in Figure~\ref{fig:modularity}, which illustrates a non-modular (a) and a modular (c) data sets as well as the respective dendrograms (b) and (d).

Henceforth, the term \emph{hierarchical modularity} is understood to mean modularity quantified along a scale variable, being associated to the hierarchy of groups. The hierarchical modularity of the dataset in (c) is reflected in the presence of three relatively large and balanced (similar number of points) clusters that extend widely along the scale variable $s$ in the dendrogram shown in (d). At the same time, less modular dendrogram in (b) is characterized by smaller and unbalanced clusters extending irregularly along the scale variable. Interestingly, these three main properties involved in the quantification of modularity are directly represented in a dendrogram.

The present work develops the possibility of, given a dendrogram, estimating its possible hierarchical cluster structure and overall modularity. This potential is illustrated by the values of overall modularity $M$ obtained by the methodology described in this work, which are also shown in Figure~\ref{fig:modularity}. Although the approach focuses on hierarchical clustering structures, the described concepts and methods can also be used in cases characterized by a single hierarchical level.

The reported approach is based on two main guidelines: (a) the consideration of the \emph{merging density function} characterizing the structure of given dendrogram along successive scales (s); and (b) the application of an equalizing approach in which branches at higher hierarchies are given larger weights.

Once an equalized merging density function has been obtained from a dendrogram within a specified resolution, its valleys are detected, allowing the main branching levels to be identified. An estimation of the hierarchical modularity can also be obtained from the dendrogram. In particular, the concept of \emph{infinitesimal modularity} is presented and then used to obtain average, overall, individual, set, and groups modularity. The basic adopted principle is that, since the infinitesimal modularity quantifies the modularity at a specific scale (associated to the vertical axis of dendrograms), it becomes possible to take into account the extension of the modularity along the scale variable by integrating the infinitesimal modularity on intervals associated to a single group, a set of groups, or all the groups at a given dendrogram slice. Similarly, it is also possible to obtain average values of the infinitesimal along given intervals. Except for the overall modularity, all other types of modularity described in the present work can be considered to be \emph{local} in the sense that their values depend only on an interval along the region of interest of the variable $s$, not being influenced by the remainder of the dendrogram.

Table~\ref{tab:modularities} summarizes the considered modularities, presenting their brief description, as well as the respectively adopted symbols. The potential of these approaches is illustrated for some types of dendrograms and data sets.

\begin{table}[]
\caption{Summary of the modularities presented and used in the current work, their adopted symbols, and brief respective description.}
\label{tab:modularities}
\begin{tabular}{|c|c|l|}
\hline
\textbf{Modularity Name} & \textbf{Variable} & \multicolumn{1}{c|}{\textbf{Brief Description}} \\ \hline
infinitesimal & $u(s)$ & modularity at a specific scale $s$ \\
average       & $\left< u \right> _{[a,b]}$ & average modularity along $s \in [a,b]$ \\
overall        & $M$ & modularity of the whole dendrogram \\
individual & $\nu(A)$ & modularity along a given group $A$ \\
set           & $\gamma(G)$ & modularity along a set of groups $G$ \\
groups           & $m(s)$ & modularity of a all groups at a level $s$ \\
\hline
\end{tabular}
\end{table}

In addition to describing the concepts and methodology for estimation of the hierarchical cluster structure of a dendrogram and its hierarchical modularity, the present work also includes a preliminary analysis of how a parameter involved, namely $\tilde{\sigma}$, can influence the estimation of the main merging levels. An additional approach to identifying the hierarchical main merging levels, based only on the signature of infinitesimal modularity is also presented.

Case-examples are also included in order to illustrate the potential of the reported methodology, including the possibility to consider sub-dendrograms in recursive manner. An application to the hierarchical analysis of additive dendrograms, more specifically those obtained from two-dimensional DLA structures, is also described. At least for the considered types of dendrograms and configurations, the proposed approach allowed interesting results concerning both the identification of hierarchical clusters and the estimation of several aspects of the respective modularity.

This work starts by briefly reviewing basic related concepts from hierarchical clusters, followed by a brief review of some related works. The work then proceeds to present merging density functions, their equalization, the identification of the main merging levels, cluster estimation, and modularity approaches, which are followed by a discussion of the influence of the resolution and a possible relationship with Bayesian Decision Theory. The potential of the described concepts and methods is subsequently illustrated for some types of dendrograms and data sets. The work concludes with a summary of the main results and some possibilities for future developments.

\section{Hierarchical Agglomerative Approaches: Basic Concepts}

Hierarchical agglomerative methods (e.g.~\cite{duda2000pattern, theodoridis2006pattern}) progressively merge data elements and subclusters according to monotonically increasing values of a scale parameters $s$, which may correspond to some measurement of the distance or dissimilarity between the subclusters. Each subsequent merging, which takes place at a specific value of $s$, defines a respective branch in the associated dendrogram. Mergings can be performed considering several criteria that include, but are not limited to, single, complete, and average linkage (e.g.~\cite{duda2000pattern, theodoridis2006pattern}). The latter approach, which is considered in the present work, involves taking the average of the distance of similarity measurements among every pair of elements between two given subclusters. However, observe that the described concepts and approaches are obtained directly from the given dendrogram, so that the linkage criterion is used here only as one of the possible means of illustrating the approach respectively to dendrograms obtained from associated data sets.

\begin{figure}[h]
  \centering
     \includegraphics[width=0.9 \textwidth]{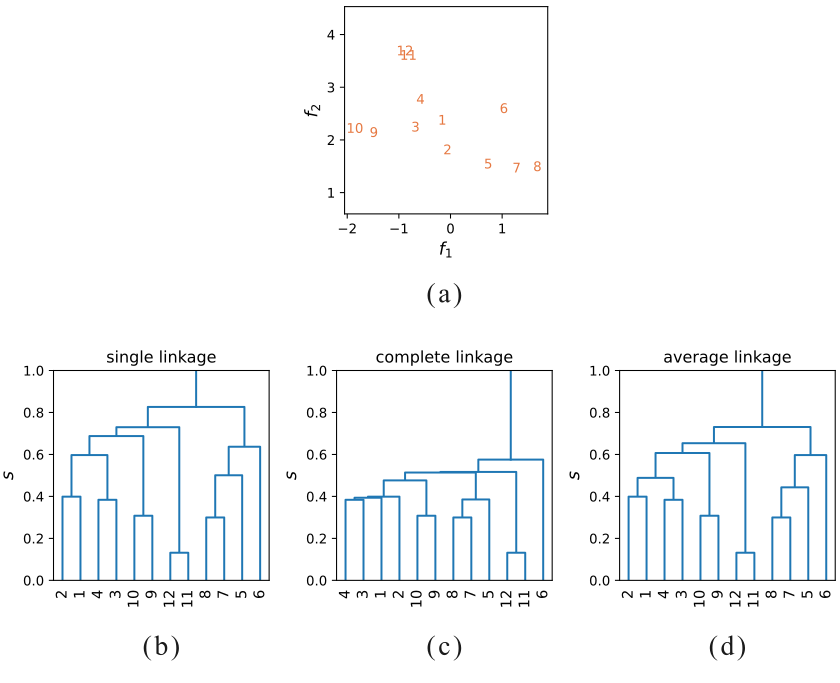}
 \caption{A simple dataset (a) and dendrograms obtained by using three distinct linkage criteria.}\label{fig:linkage}
\end{figure}

Figure~\ref{fig:linkage} illustrates a simple distribution of $12$ points in (a), considered as data elements, and the respectively obtained dendrograms considering single in (b), complete in (c), and average in (d) linkage criteria.

As can be verified in Figure~\ref{fig:linkage}, the choice of linkage criterion can lead to different dendrograms. The choice of the linkage criterion, which is not addressed further in the present work, depends on each case, type of data, and eventual presence of noise, among other factors. The separation (and therefore modularity) of the data elements encompassed by a respective branch has been related to the length of its stem, among other possibilities (e.g.~\cite{karna2022}).

The present work concentrates on cluster detection and modularity estimation directly from given dendrograms, not specifically addressing how they have been obtained. In the cases described herein in which dendrograms are obtained from data sets, the average linkage criterion has been adopted. 

The above approach reflects the fact that the task of pattern recognition can be split into three main subsequent stages: (i) the estimation of interrelationships between the data elements (e.g.~the Euclidean distances between the elements features) yielding a respective weight matrix; (ii) transformation of the weight matrix into a dendrogram; and (iii) analysis of the dendrogram in order to find hierarchical clusters and estimate their modularity. Despite such a division of tasks, the choices and results obtained at each level potentially still impact the other stages. At the same time, focusing on one of the stages involved allows special attention and efforts to be concentrated on the respective concepts and methods. The present work focuses on (iii).

\section{Related Works}

The subjects of hierarchical clustering, modularity, and dendrograms have been the subject of several related works. The present section briefly reviews some of these methods, especially from the perspective of hierarchical analysis of dendrograms and hierarchy of DLA structures.

The identification of the main clusters in hierarchical structures, including dendrograms has been addressed by using several ways of assigning some relevance measurement to how subsets of data elements are separated from the other elements in the dendrogram. These approaches are often based on the two following aspects: (i) length of the branches along the scale variable; and (ii) homogeneity of distribution of elements among the branches cut at a specific value of the scale variable.

The identification of relevant merging levels in terms of variation of the scale variable values where they take place has been addressed in~\cite{mojena} in terms of two stopping rules, based on the distribution of the merging levels. An approach to estimate the significance of hierarchical clusters, implemented as a sequence of tests, has been described in~\cite{kimes2017} which considers Monte Carlos simulations. The identification of clusters in hierarchical trees has been addressed in~\cite{langfelder2008} which uses a dynamic branch cutting procedure while taking into account their shapes as well as the stability of the clusters. The approach is complemented by a bottom-up method for improving the identification of outlier elements. An approach for slicing dendrograms based on the evaluation of the need to further subdivide clusters~\cite{palazzo2026} involves the application of a top-down permutation-based test procedure. The significance of branches in hierarchical trees has been approached~\cite{sun2014} from the perspective of ratios between heights of both a branch and its parent. An approach has been described~\cite{karna2022} for estimating the number of clusters in hierarchical clustering combining several criteria using the Calinski-Harabasz index~\cite{calinski1974} (ratio between clusters dispersions).

The asymmetry of partitioning of clustering trees has been quantified by using entropy in~\cite{corredor2020}. The partitioning of dendrograms was addressed in terms of entropy parameters in~\cite{casado1997}. More specifically, the homogeneity of distribution of the species into groups is quantified in terms of the respective entropy, which is normalized between $0$ and $1$ (main niche width). The main merging level is identified by minimizing this normalized entropy. The quantification of the flatness of distance distributions in a dataset in terms of normalized entropy has been addressed in~\cite{metzner2023}, indicating that well-separated clusters tend to be associated with large peaks of the distance distribution. The quantification and optimization of the relevance of dendrograms merging levels in terms of a normalized entropy has been considered in~\cite{martinezarmas2025}, involving recursive application of the entropy criterion according to a renormalized framework.

The aggregation of spherical conducting particles in a dielectric liquid medium tends to yield trees which have had several of their topological and geometrical properties studied in~\cite{soni2011}, including bifurcation and lengths ratios, which were found to agree with those obtained in respective DLA simulations and to have similarities with hydrological networks. Bifurcation and lengths ratios have also been considered for the study of dendrograms obtained from axons in the cat visual cortex~\cite{binzegger2005}. Neuronal structures have also been studied from the perspective of DLA modeling~\cite{luczak2006}.

\section{Merging Density Functions}\label{sec:DensityFunction}

Given a dendrogram with $N$ elements along a scale parameter $s \in S$, it is possible to consider its merging levels:
\begin{align}
L=\{s_1,s_2,\dots,s_{N-1}\},
\qquad s_i \in S,
\end{align}
where $s_i$ denotes the height of the $i$-th merging event and $N-1$ is the number of merging levels. Observe that $s_i < s_{i+1}$ for $i=1, 2, \, \dots \, N-2$. 

Consider the following function $f(s)$, which corresponds to the sum of Dirac's deltas at each merging scale $s_i$, as a representation of the distribution of merging events along the hierarchy:
\begin{align}
  f(s) = \frac{1}{N-1} \sum_{i=1}^{N-1} \, \delta(s_i)
\end{align}

A smoothed version $q(s)$ of the discrete function above can be obtained by convolving $f(s)$ with a Gaussian function (e.g.~\cite{brigham1988fft}) with fixed standard deviation $\tilde{\sigma}$, corresponding to a parameter that establishes the \emph{resolution} of the merging density:
\begin{align}
  q(s) = \frac{1}{N-1}\sum_{i=1}^{N-1} \, \frac{1}{\tilde{\sigma}\sqrt{2\pi}} \, \exp\left(-\frac{(\tilde{\sigma}-s_i)^2}{2\tilde{\sigma}^2}\right).
\end{align}

Each merging level can be associated with a weight $w_i$, proportional to the number of elements covered by the branch with merging level $s_i$, leading to a new discrete weighted function:
\begin{align}
  \tilde{f}(s) = \frac{1}{N-1} \sum_{i=1}^{N-1} \, w_i\,\delta(s_i).
\end{align}

In addition to the number of elements covered by the branch, it is also possible to consider the geometric or harmonic averages, or the maximum and minimum (among other possibilities) of the number of elements among the sub-branches. In order to avoid accumulation of mergings involving one (or two) small branches (as in single linkage chaining, e.g.~\cite{tokuda2022}), it is also possible to take into account a threshold for the mergings being considered, therefore only including mergings between subgroups which have at least a minimum number of elements. Weights can also be assigned to specific data elements (leaves) as a possible means to control the relevance of respective elements, in the case of specific application requirements.

The \emph{equalized merging density} $p(s)$ is herein defined as a smoothed version of $\tilde{f}(s)$, indicated in the following:
\begin{align}
r(s) &=\sum_{i=1}^{N-1} w_i \, \exp\left(-\frac{(\tilde{\sigma}-s_i)^2}{2\tilde{\sigma}^2}\right),
\\
p(s) &= \frac{1}{\int_S r(s)\, ds}\; r(s).
\end{align}

In addition to linear smoothing by convolution with Gaussians, it is also of potential interest to consider non-linear smoothing approaches (e.g.~\cite{perona1994,black1998,weickert1998}) capable of preserving more effectively the singularities (peaks) along the equalized merging density $p(s)$. 

\begin{figure}[h]
  \centering
     \includegraphics[width=0.9 \textwidth]{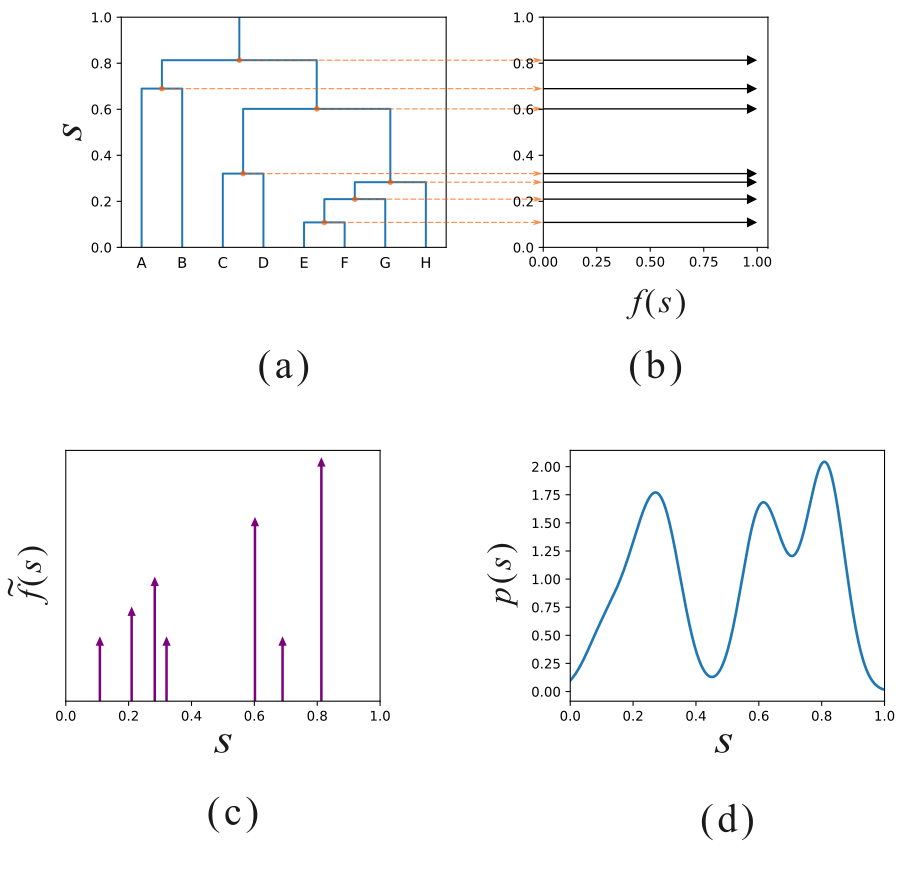}
 \caption{Illustration of the estimation of the functions $f(s)$ (b), $\tilde{f}(s)$ (c), and $p(s)$ (d) respectively to the dendrogram in (a). These results have been obtained for $\tilde{\sigma}=0.06$.}\label{fig:density}
\end{figure}

Figure~\ref{fig:density} illustrates the approach to obtain the equalized merging density $p(s)$ from a given dendrogram (a).

\begin{figure}[h]
  \centering
     \includegraphics[width=0.7 \textwidth]{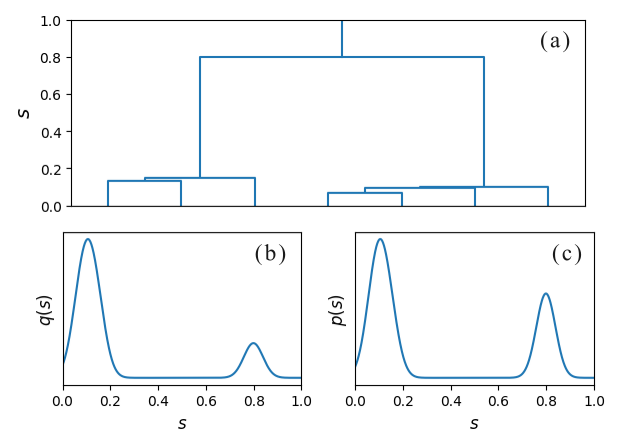}
 \caption{A dendrogram with two main hierarchical levels is shown in (a). Its original and equalized merging density function is presented in (b) and (c), respectively. The effect of the adopted equalization is to enhance the levels at higher hierarchies.}\label{fig:balancing}
\end{figure}

The adoption of the equalized version of the merging density function is motivated as follows, considering the dendrogram shown in Figure~\ref{fig:balancing}(a). The associated merging density function ($q(s)$) is shown in Figure~\ref{fig:balancing}(b), which presents two main hierarchical levels (at $s \approx 0.10$ and $s \approx 0.80)$. The equalized merging density ($p(s)$) is shown in Figure~\ref{fig:balancing}(c). As expected, the peak at the high hierarchy (large value of $s$) has been enhanced. 

This result illustrates that main levels at higher hierarchies tend to result in relatively low peaks in the merging density function. In order to enhance these peaks, the \emph{equalizing approach} described above has been adopted henceforth.

\section{Hierarchical Clusters Estimation}\label{sec:cluster}

Given a dendrogram and its respective equalized merging density function $p(s)$, it is interesting to try to estimate clusters eventually present in the original data set from which the dendrogram was obtained. A possible non-supervised way to do so is described in this section.

\begin{figure}[h]
  \centering
     \includegraphics[width=0.9 \textwidth]{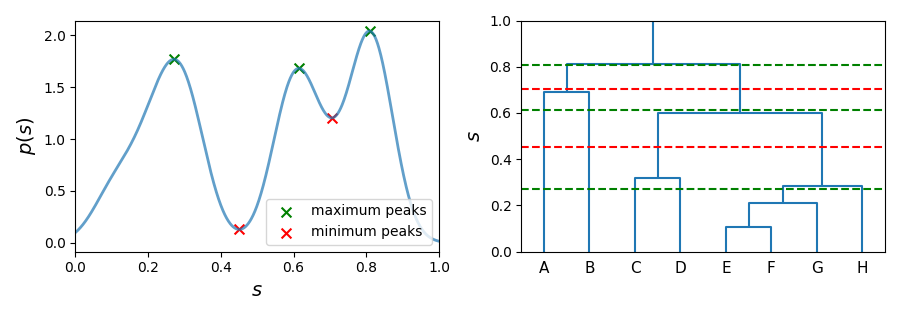}\\
     \hspace{1 cm} (a) \hspace{5 cm} (b)\\

    \vspace{0.5 cm}
     \includegraphics[width=0.45 \textwidth]{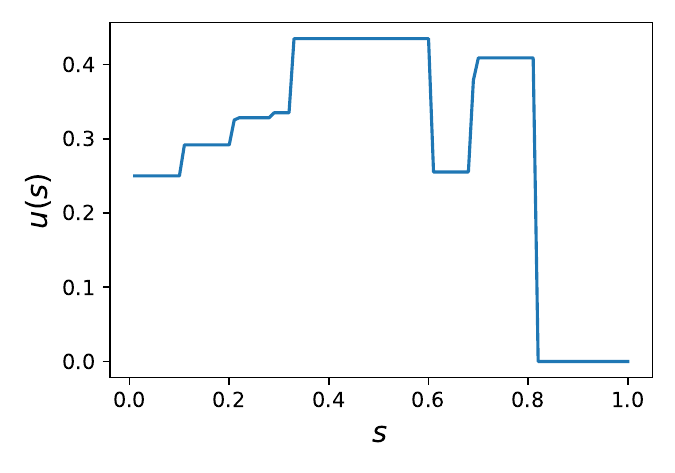}\\
     \hspace{1 cm} (c)
     
 \caption{Example of detected maximum and minimum peaks of $p(s)$ in (a). The dendrogram and the respective merging levels defined by these peaks are presented in (b). The respectively obtained signature of infinitesimal modularity is shown in (c).}\label{fig:valleys_peaks}
\end{figure}

First, maximum peaks along $p(s)$ tend to be associated with the \emph{main branching levels} of a dendrogram. Figure~\ref{fig:valleys_peaks}(a) shows this type of level identified for the function $p(s)$ associated with the dendrogram in Figure~\ref{fig:valleys_peaks}(b).

Henceforth, the minimum peaks, together with the two limits of the scale variable $s$, are called the \emph{delimiting levels} of a dendrogram. The sequence of delimiting levels defines $n$ subsequent intervals, $I_k = [v_k, v_{k+1}]$, $k= 1, 2, \ldots, n$, with $v_0 = 0$ and $v_n = 1$, on the scale $s$. For example, the intervals identified for the dendrogram in Figure~\ref{fig:valleys_peaks} are $I_1 = [0.0,0.45]$, $I_2 = [0.45,0.71]$ and $I_3 = [0.71,1.0]$.

Although other implementations and approaches are possible, in the present work the maximum and minimum peaks of $p(s)$ are identified by applying the \texttt{find\_peaks} algorithm to $-p(s)$ using the SciPy signal-processing package \cite{scipy_findpeaks}. This algorithm identifies local maxima in a one-dimensional signal by comparing neighboring values and applying optional constraints such as prominence, height, width, and minimum distance between peaks. The minimum peaks in $p(s)$, corresponding to the valleys of that density, can also be identified by applying the same algorithm to $-p(s)$. 

Figure~\ref{fig:valleys_peaks}(a) also illustrates minimum peaks identified along the function $p(s)$ obtained for the dendrogram in Figure~\ref{fig:valleys_peaks}(b). The levels corresponding to the detected maximum and minimum peaks are also superimposed on the dendrogram in that figure.

The complete signature of infinitesimal modularity obtained for the dendrogram in Figure~\ref{fig:valleys_peaks}(c) indicates how this measurement changes along the hierarchical structure of the dendrogram. More specifically, after an initial increase up to the maximum observed modularity, the signature presents an isolated peak and then falls to 0 (single group observed at the end of the $s$ scale). The first decrease was implied by the merging of two less balanced groups.

Although the positive and negative peaks of $p(s)$ provide important information about the possible presence of clusters in the original data, they are not enough to identify those clusters because more than one group may be associated to the same delimiting level. Therefore, a subsequent analysis is required that involves the identification of the data elements comprehended by the branches cut by each of the delimiting levels.

\begin{figure}[h]
  \centering
     \includegraphics[width=0.9 \textwidth]{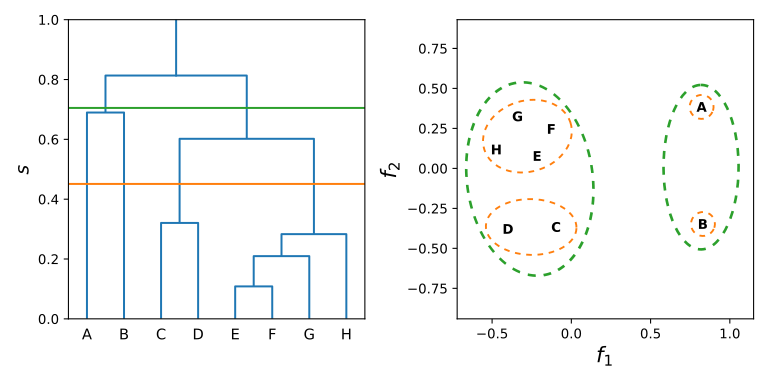}\\
     \hspace{.8 cm} (a) \hspace{5.2 cm} (b)
 \caption{Illustration of cluster detection for the dendrogram in (a). The two identified delimiting levels are shown in orange and green in (a), and the respectively detected hierarchical group organization are presented in (b).}\label{fig:density_clustering}
\end{figure}

Figure~\ref{fig:density_clustering}(b) presents the hierarchical clusters which have been obtained for the dendrogram in Figure~\ref{fig:density_clustering}(a) by using the described methodology with $\tilde{\sigma}=0.06$.

\section{Infinitesimal, Average, Overall, Individual, Set, and Groups Modularities}

Given a dendrogram with $N$ data elements, it is possible to consider its \emph{infinitesimal modularity} at any specific value of $\tilde{s}$ of the scale variable $s$ while taking into account only the number of data elements $N_i$ contained in each of the clusters $i=1, 2, \ldots,c$ indicated by the respective slice. Observe that $N= N_1+N_2+\ldots+N_c$. In this work, the infinitesimal modularity at $\tilde{s}$ is henceforth understood as:
\begin{align}\label{eq:infinitesimal_modularity}
    u(\tilde{s}) = & \left[ \frac{\min\left(c_e \, \tilde{h}, h \right)}{\max\left(c_e \, \tilde{h}, h \right)} \right]^D \, \left[ \frac{\min \left(2, c_e \right)}{\max \left(2, c_e \right)} \right]
\end{align}
where $h(N_1,N_2,\ldots,N_c)$ is the entropy function and $c_e = \exp \left(h(N_1,N_2,\ldots,N_c) \right)$ is the effective number of clusters, i.e.:
\begin{align}
    c_e = \exp(h) = \exp \left( -\sum_{i=1}^c\frac{N_i}{N} \, \log\left(\frac{N_i}{N} \right) \right),
\end{align}
$\tilde{h}(N_1,N_2,\ldots,N_c)$ is the weighted entropy function, and $D$ is a positive exponent that controls how strict the balance between the number of elements in the branches is to be implemented. More specifically, the larger the value of $D$, the more strict the balance quantification becomes.

The weighted entropy function (e.g.~\cite{guiacsu1971}) $\tilde{h}(N_1,N_2,\ldots,N_c)$ in Equation~\ref{eq:infinitesimal_modularity} can be expressed as:
\begin{align}\label{eq:balanced_entropy}
    \tilde{h} \left( N_1,N_2,\ldots,N_c \right) = - \sum_{i=1}^c \, w_i \, p_i \, \log(p_i) 
\end{align}
where $p_i$, $i=1, 2, \ldots, c$ are probabilities adding to one and $w_i$ is a weigh associated to cluster $i$. In the present work, we have $p_i = N_i/N$ and $w_i = p_i$, so that the probabilities are weighted by the number of data elements in the respective cluster.

Observe that the two main terms in Equation~\ref{eq:infinitesimal_modularity} are normalized in the interval $[0,1]$, so that the resulting infinitesimal modularity is also bound within $[0,1]$.

Equation~\ref{eq:infinitesimal_modularity} assumes that the infinitesimal modularity $u(\tilde{s})$ of a given dendrogram depends of the balance between the number of data elements in each cluster, which is quantified by the first term in that equation. We have that the balance between the number of data elements in the identified clusters reaches its maximum value 1 whenever $N_1=N_2=\ldots=N_c$. It is interesting to observe that the balance among clusters also reflects the inner dispersion of any of the involved clusters since it will tend both to reduce the shortest distance to the other clusters but also to potentially decrease the balance when considered among other clusters. In addition, well balanced clusters will tend to present not only relatively high, but also similar infinitesimal modularity values. 

Because the same balance value can be obtained for any number of clusters $c$, with $c \geq 2$, it is interesting to weight the entropy linearly by the effective number of clusters, which is implemented in terms of the exponential of the entropy in the second term of Equation~\ref{eq:infinitesimal_modularity}. Consequently, the infinitesimal modularity will increase in cases where the numbers of data elements in the $c$ clusters are effectively larger.

All in all, the above definition of infinitesimal modularity reflects the \emph{evenness of distribution of the data elements among the clusters} (quantified by the first term) as well as the \emph{sizes of these clusters} (the second term in the equation). Therefore, given $N$ data elements, the maximum infinitesimal modularity value of 1 will be obtained provided that $N$ is even and only two branches are involved, each having precisely $N/2$ data elements.

As a numeric example, in the case of the dendrogram in Figure~\ref{fig:density}(a), the infinitesimal modularity $u$ on the slice $\tilde{s} = 0.4$ for $D=3$ can be obtained as follows:
\begin{align}
   &c = 4;  \nonumber \\
   &N_1 = 1; \; N_2 = 1; \; N_3 = 2; \; N_4 = 4; \nonumber \\
   &N = 8;  \nonumber \\
   &h \approx 1.213008;  \nonumber \\
   &c_e \approx 3.363587;  \nonumber \\
   &\tilde{h} = 0.324913;  \nonumber \\
   &c_e \, \tilde{h} = 1.092872;  \nonumber \\
   &u(\tilde{s}) = u(0.4) =  \left[ \frac{\min\left(c_e \, \tilde{h}, h \right)}{\max\left(c_e \, \tilde{h}, h \right)}  \right]^3  \; \left[ \frac{\min \left(2, c_e \right)}{\max \left(2, c_e \right)}  \right] \approx \nonumber \\
   &\approx \left[ \frac{1.092872}{1.213008}  \right]^3  \; \left[ \frac{2}{3.363587}  \right] = \left[ 0.731336 \right] \, \left[ 0.5946036 \right] = 0.434855
\end{align}

Given that infinitesimal modularity is a continuous (but not differentiable) function of $s$ that can be estimated for the whole considered interval of the scale variable $s \in [0,1]$ (adaptations to other ranges are also possible), it becomes possible to integrate this function along specific intervals of interest. In particular, in this work we are interested in the \emph{average modularity} and \emph{overall modularity} of a dendrogram, which are defined in the following.

The \emph{average modularity} of a dendrogram along an interval $[a,b]$ can now be obtained as:
\begin{align}
    \left< u \right> _{[a,b]} = \frac{1}{b-a} \int_{a} ^{b} \, u(s) \, ds
\end{align}

As an example, in the case of the dendrogram in Figure~\ref{fig:density}(a), the average modularity for $s \in [0.2,0.6]$ is $v(0.2,0.6) = 0.402$.

In particular, when $a=0$ and $b=1$, we get the \emph{overall hierarchical modularity} $M$, which provides an indication of the global modularity of the complete given dendrogram. In other words, we have:
\begin{align}
    M = v(0,1) = \int_{0} ^{1} \, u(s) \, ds
\end{align}

For example, the dendrogram in Figure~\ref{fig:density}(a) has overall hierarchical modularity $M=0.389$.

Another useful concept relates to the modularity of a \emph{single group} $A$ in a given dendrogram. For simplicity's sake, groups and subgroups are henceforth referred to \emph{groups}. Recall that a group in a dendrogram can be specified by its respective stem or by the set of data elements which are included in that group. Given that the modularity of a single group taken separately would be zero (the quantification of modularity needs a reference), it is necessary to consider additional elements from the original data. At the same time, the modularity of the given group $A$ is a \emph{local property}, in the sense of not (or little) being affected by data elements that are further away. 

In the present work, the locality of the group $A$ is taken as corresponding to all subgroups that are at the maximum value of the scale variable where the group $A$ first merges with another group $Z$. Therefore, the \emph{individual modularity} of a group $A$ can be quantified as:
\begin{align}
    \nu(A) = \int_{A_a} ^{A_b} \, \tilde{u}(s) \, ds 
\end{align}
where $\tilde{u}(s)$ is the infinitesimal modularity function obtained only for the subdendrogram formed by all the elements of $A$ plus the elements of the group $Z$ into which it merges, and $A_a$ and $B_b$ correspond to the smallest and largest values of $s$ along the stem of the group $A$, respectively.

Now, it becomes possible to consider the modularity of a set $G$ of groups of a given dendrogram. This can be done by selecting the subdendrogram containing all the elements of each of those groups, plus the elements of the groups into which they merge for the first time. The respective infinitesimal modularity function $\tilde{u}$ can then be obtained, and the \emph{set modularity} of $G$ can be calculated as:
\begin{align}
    \gamma(G) = \int_{G_a} ^{G_b} \, \tilde{u}(s) \, ds 
\end{align}
where $G_a$ and $G_b$ are the lower and upper limits of the interval of $s$ defined by the \emph{intersection} of the stems of all the groups that belong to $G$. The intersection is taken because the set modularity refers to the \emph{co-existence} of all considered groups along the scale variable.

It is also interesting to consider the modularity of \emph{all the groups} identified by a respective slice $\tilde{s}$, which is henceforth called the \emph{groups modularity}. This can be obtained by calculating the set modularity for $G$ containing all groups identified by the respective slice. Therefore, we have:
\begin{align}
    m(\tilde{s}) = \int_{\alpha(\tilde{s})}^{\beta(\tilde{s})} \, u(s) \, ds
\end{align}
where:
\begin{align}
    \alpha(\tilde{s}) = \max_{i=1}^c \left\{ s_{s,i} \right\} \nonumber \\
    \beta(\tilde{s}) = \min_{i=1}^c \left\{ s_{f,i} \right\} \nonumber 
\end{align}
where $s_{s,i}$ and $s_{f,i}$ are the starting and ending points of the group $i$ along the variable $s$, for $i = 1, 2, \ldots, c$.

\section{Extension to Additive Dendrograms}

So far in this work, we have considered dendrograms whose all leaves start at the scale variable $s=0$. That property is a consequence of the fact that each separate data element has distance 0 (or similarity 1) at $s=0$. These types of structures, characterized by all leaves having the same distance to the root, are called \emph{ultrametric trees}~\cite{corter1996,sattath1977}.

However, there are several cases in which the dendrograms have a common root, but the leaves are not at the same height, as happens in phylogenetic trees, language evolution, and historic or risk maps, among other situations. These trees can be understood as being \emph{non-ultrametric}. Frequently, this type of dendrogram has scale variable which can not be associated with time, arc-length distance, number of changes, or other cumulative measurements. Other examples of this type of structure are branched geometric structures that include roots, river systems, blood vessels, and bronchial tubes, among many others. This type of structure can be understood as \emph{additive trees} (or \emph{dendrograms}), which also incorporate structures based on additive similarity~\cite{corter1996,sattath1977} aimed at minimizing the variance of the distance between the root and leaves.

\begin{figure}[h]
  \centering
     \includegraphics[width=0.99 \textwidth]{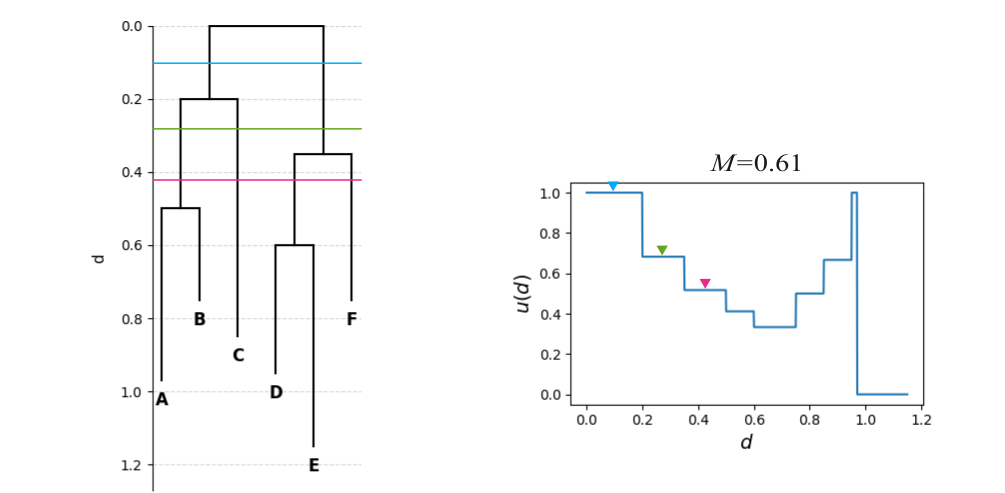}\\
     (a) \hspace{6 cm} (b) \\ \vspace{0.55 cm} 
     \includegraphics[width=0.5 \textwidth]{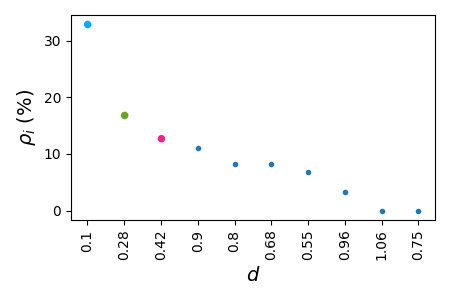}\\
      \hspace{0.5 cm} (c) \\
  \caption{A simple additive dendrogram (a), its infinitesimal modularity signature (b), sorted relevance values of modularity (c). The three main identified merging levels are also shown in (a).} \label{fig:ex_addtree} 
\end{figure}

Interestingly, the same concepts and methods described above in the present work can be applied directly to additive dendrograms. As an example, consider the simple additive dendrogram in Figure~\ref{fig:ex_addtree}(a) considering an absolute scale of distances (not normalized to relative values).

The infinitesimal modularity signature $u(d)$ of the additive dendrogram in terms of the distance is shown in (b), and the sorted values of relevance modularity are presented in (c).

In the above example, the infinitesimal modularity took into account only the data elements contained in the groups cut at the given height of the dendrogram. Another possibility consists of using the total number of data elements irrespectively of the cut. The choice among these, as well as other possibilities, will depend on the specific type of data and/or interest of each research situation.

\section{Influence of the Parameter $\tilde{\sigma}$}\label{sec:sigParameter}

The concepts and methods presented above involve one main parameter, namely the fixed smoothing scale $\tilde{\sigma}$ corresponding to the standard deviation of the Gaussian function used to interpolate and smooth the function $\tilde{f}(s)$.

As indicated in Section~\ref{sec:DensityFunction}, the parameter $\tilde{\sigma}$ specifies the \emph{resolution} of cluster detection. Smaller values of $\tilde{\sigma}$ will allow the finer structure of a dendrogram (branches that are closer to each other) to be taken into account, which tends to yield a larger number of main levels and clusters. At the same time, larger values of $\tilde{\sigma}$ will allow the approach to emphasize the larger scale structure of a dendrogram, merging the branches which are at scales closer to each other. The choice of $\tilde{\sigma}$ depends on several aspects, including the main objectives of the analysis, the structure of the data, and the eventual presence of noise or other types of interference in the original data.

Important related issues concern the possible effects of different choices of $\tilde{\sigma}$ on the obtained clusters. These aspects are addressed in the present section.

\begin{figure}[h]
  \centering
     \includegraphics[width=0.99 \textwidth]{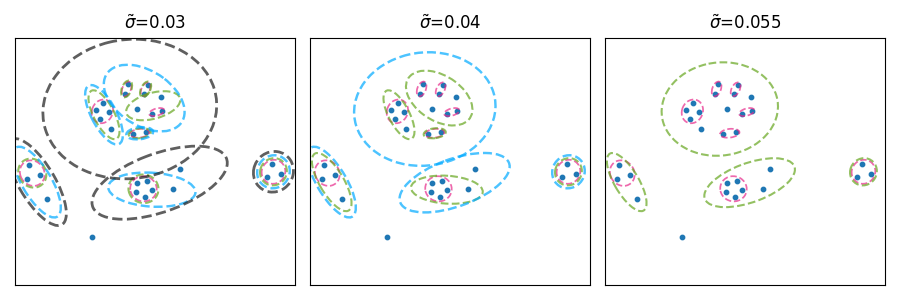}\\
     (a) \hspace{3.5 cm} (b) \hspace{3.5 cm} (c)
 \caption{The hierarchical clusters obtained for the dendrogram in Figure~\ref{fig:case_den2} by using $\tilde{\sigma}=0.03, 0.04,$ and $0.055$. As could be expected, less detailed (lower resolution) clusters are obtained as $\tilde{\sigma}$ increases.}\label{fig:group_s} 
\end{figure}

Figure~\ref{fig:group_s} illustrates the identified delimiting levels and associated detected clusters for the same dendrogram, but using three increasing values of $\tilde{\sigma}$.

These results indicate that, at least for the case considered in this example, relatively small changes in the hierarchical clustering structure were implied even by large variations of the parameter $\tilde{\sigma}$.

\section{Relationship with Bayesian Decision Theory}

This section addresses the relationship between the concepts and approach described above and Bayesian Decision Theory (e.g.~\cite{berger1985, duda2000pattern, theodoridis2006pattern}), which consists of an optimal approach to \emph{supervised} classification. This relationship ultimately provides a motivation and justification for the present work.

We start with a brief review of Bayesian Decision Theory in an one-dimensional feature space $x$. Given $M$ groups of elements corresponding to classes $C_k$, each of them exactly described by a respective density function $p_k(x)$ and mass probabilities $P_k$, the chances of misclassification of a new sample with feature $\tilde{x}$ can be optimally minimized by using the following criterion:
\begin{align}
\text{chosen class} = \, \tilde{k} \; \Big| \; P_{\tilde{k}} \, p_{\tilde{k}}(\tilde{x}) = \max_k \left\{ P_k\, p_k(\tilde{x}) \right\}. \label{eq:Bayes}
\end{align}

In practice, the above criterion implies that the boundaries between the categories along the feature variable $x$ will correspond to the points of intersection between respective density functions $p_k(x)$.

In the case of the approach described in the present work, assume that each of the groups of merging levels $k = 1, 2, \ldots, n$ are described by exact respective density functions $p_k(s)$, therefore corresponding to a supervised classification problem and allowing Bayesian Decision Theory to be respectively applied in order to identify the main hierarchical levels.

As an approximate method, it would be possible to consider the function
\begin{align}
    \eta(s) = \sum_k P_k \, p_k(s).
\end{align}
Provided the groups are relatively well-separated, the minimum peaks of $p(s)$ would approximate the intersections obtained by applying Bayesian Decision Theory considering the separated densities as indicated in Equation~\ref{eq:Bayes}.

Because the present work addresses unsupervised classification (identification
of the hierarchical levels), the function $\eta(s)$ cannot be known a priori, as this would require knowledge about the groups. What is available instead is the function $p(s)$, which can be related to the separate density as:
\begin{align}
    p(s) = \sum_k \, p_k(s)
\end{align}

In the case of relatively well separated groups, the minimum peaks identified in $p(s)$ tend to approximate those along $\eta(s)$. Under these circumstances, it becomes possible to obtain a potentially improved version of $\eta(s)$ by approximating $p_k(s)$ by non-parametric estimated densities $\tilde{p}_k$ within each detected interval as follows:
\begin{align}
    \tilde{\eta}(s) = \sum_k P_k \, \tilde{p}_k(s)
\end{align}
where $P_k$ is the number of merging levels within each respective interval.

\begin{figure}[h]
  \centering
     \includegraphics[width=0.48 \textwidth]{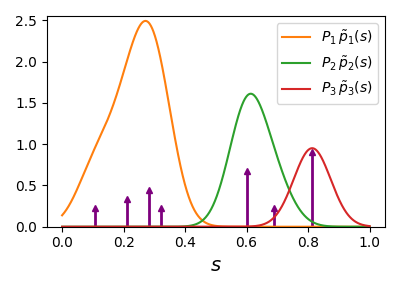}
     \includegraphics[width=0.48 \textwidth]{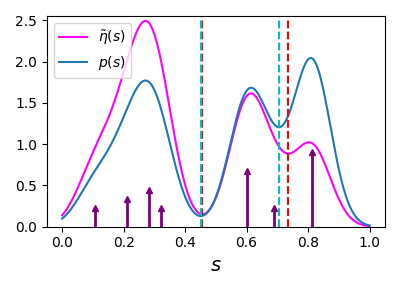}\\
     \hspace{1.1 cm} (a) \hspace{4.9 cm} (b)
 \caption{The approximated densities $\tilde{p}_k$, weighted by $P_k$, obtained by Parzen windows non-parametric estimation (a). The functions $\eta(s)$ and $p(s)$ respective to the case shown in Fig.~\ref{fig:valleys_peaks} are shown in (b) together with the identification of the respective minimum peaks (dashed lines). Observe that the separation margin between the groups of mergings obtained from $\tilde{\eta}(s)$ (dashed line in red) has been improved thanks to the weighting by the mass probability.}\label{fig:bayesian}
\end{figure}

Figure~\ref{fig:bayesian} illustrates the above discussion respectively to the case shown in Figure~\ref{fig:valleys_peaks}, which involves three groups (main hierarchical levels).

Each of the approximate densities $\tilde{p}_k(s)$, estimated non-parametrically by using the Parzen windows approach (e.g.~\cite{parzen1962}) using Gaussian kernel with $\sigma=0.06$, are shown in orange, green, and red in Figure~\ref{fig:bayesian}(a). The functions $\tilde{\eta}(s)$ and $p(s)$ are shown in Figure~\ref{fig:bayesian}(b) together with vertical dashed lines indicating the respective minimum peaks.

A first interesting aspect to be observed in Figure~\ref{fig:bayesian}(a) concerns the correct identification of the three groups of merging levels while considering the intersections between the estimated densities weighted by the number of elements in each group. As shown in Figure~\ref{fig:bayesian}(b), the positions of the minimum peaks in functions $\eta(s)$ and $p(s)$ are similar not only among themselves, but are also compatible with the positions of the intersections identified in Figure~\ref{fig:bayesian}(a). 

The above discussion indicates that, provided the main hierarchical levels are relatively well-separated, the described approach is related to Bayesian Decision Theory in the sense that the intervals identified by the minimum peaks tend to approximate the intersections between the separated densities associated to each of the groups. However, because the described approach is non-supervised and the separate densities are not available, its performance is not optimal, as would otherwise be the case with Bayesian Decision Theory. Nevertheless, it would still be possible to estimate the separate densities of the main hierarchical levels identified by the minimum peaks of $p(s)$, leading to a potentially improved identification of the intervals associated to each main hierarchical level detected.

\section{Detecting Main Merging Levels from Infinitesimal Modularity Signatures}

In addition to using the method described above to identify the possible main merging levels in a dendrogram (and therefore estimate the present clusters), it is also possible to consider the infinitesimal function (or signature) as the basis from which to perform this same task. More specifically, as illustrated in Figure~\ref{fig:u_modularityxS}, the main merging levels in a dendrogram tend to correspond to regions of relatively large constant value of $u(s)$ that extend widely along $s$.

Because in this work we have assumed that the dendrograms have topology of binary trees, the total of mergings will be equal to the number of data elements in the original data set, and the same number of merging levels will be identified by the above approach. Therefore, it becomes necessary to rank the relevance of each of the constant regions detected along $u(s)$. This can be implemented by recalling that the area $A_i$ underneath each of the constant value intervals $i=1,2, \ldots, N$ of $u(s)$ corresponds to the groups modularity of the dendrogram taken at any level $\tilde{s}$ comprised in the respective range of $s$ within the interval associated to each constant region. More specifically, the following measurement of the relevance of each merging level is henceforth quantified as:
\begin{align}
    \rho_i = 100 \, \frac{A_i}{A_T} \, \% = 100 \, \frac{A_i}{M} \, \%
\end{align}
where $A_T$ is the total area of the infinitesimal modularity along the whole dendrogram, which is equal to the overall modularity. Therefore, the value of $\rho_i$ of a given merging level is equal to its percentage of contribution to the overall modularity of the dendrogram.

The three triangular marks in Figure~\ref{fig:u_modularityxS} show the three most relevant main merging levels as identified by the above described approach respectively to the dendrogram in Figure~\ref{fig:case_den2} which, in this specific case, fully coincide with the main merging levels identified by the approach based on $p(s)$.

Compared to the methodology described in Section~\ref{sec:cluster}, the identification of the main merging levels based on the infinitesimal modularity is directly integrated to the modularity approach adopted in the present work and does not require any parameter (the previously described method requires the resolution parameter $\sigma$). However, at least as described here, the approach based on detecting the infinitesimal modularity constant intervals will not be able to simplify the obtained main levels by joining levels which are close one another (as performed by the Gaussian smoothing).

\begin{figure}[h]
\centering
     \includegraphics[width=0.85 \textwidth]{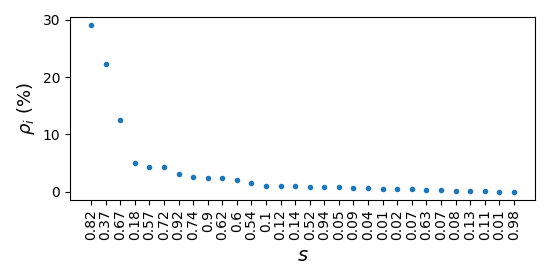}\\
 \caption{The sorted values of infinitesimal modularity of each of the merging levels obtained for the dendrogram in Figure~\ref{fig:case_den2} by using the respective infinitesimal modularity signature. The indicated values of $s$ correspond to the middle of each detected constant modularity interval.}\label{fig:sorted_levels}
\end{figure}

Figure~\ref{fig:sorted_levels} illustrates the sorted (in decreasing order) infinitesimal modularity at any of the values of $s$ within each of the $30$ identified merging levels. Interestingly, the three main levels previously identified by the approach based on Gaussian smoothing have substantially larger values of infinitesimal modularity than the remainder merging levels.

Interestingly, the developments described above indicate that dendrograms with relatively high overall modularity tend to have only a few relevant main merging levels. On the other hand, an equally chained dendrogram will yield respective merging levels decreasing linearly, therefore not highlighting any subset of levels with higher modularity.

\section{Case-Examples}

In order to illustrate the potential of the reported concepts and methods, case-examples are presented and discussed in this section.

\begin{figure}[h]
  \centering
     \includegraphics[width=0.99 \textwidth]{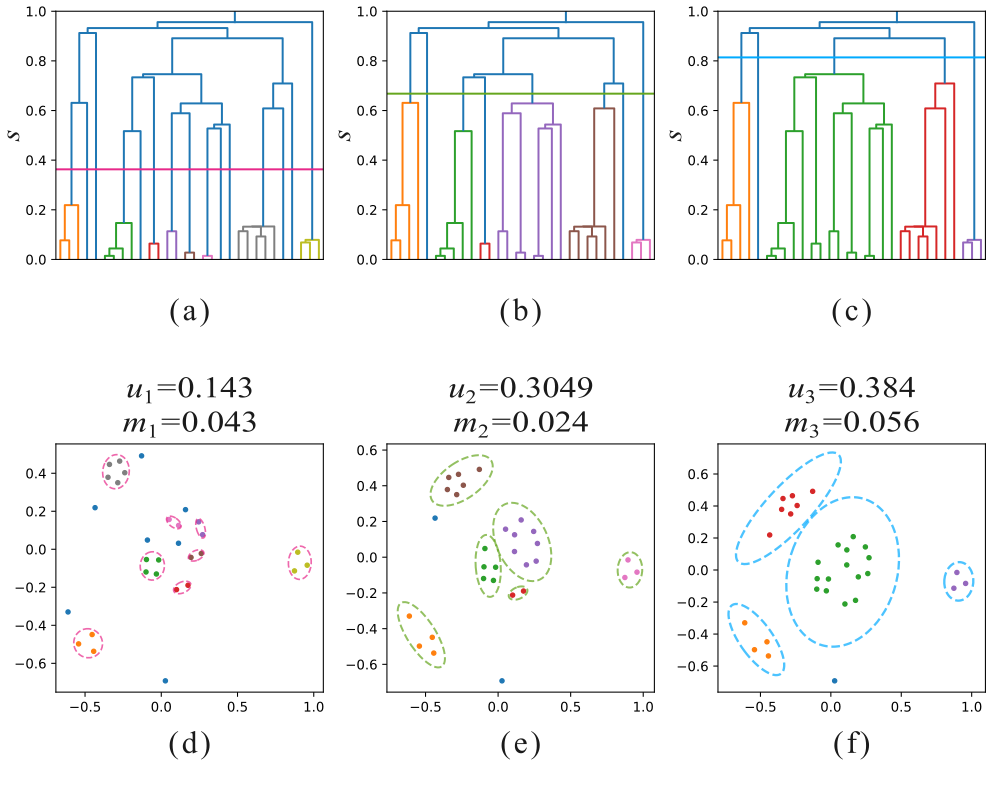}
 \caption{Illustration of clusters hierarchically detected from a dendrogram at three increasing delimiting levels. Each of these three cases are shown as columns in the figure, each presenting the respective delimiting levels (a,b,c) as well as the associated detected clusters (d,e,f). The group and infinitesimal modularities are also shown in the figure, respectively to each detected main merging levels.}\label{fig:case_den1}
\end{figure}

\begin{figure}[h]
  \centering
     \includegraphics[width=0.8 \textwidth]{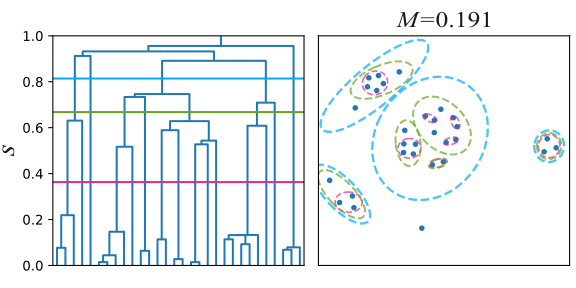}\\
     \hspace{.7 cm} (a) \hspace{4.2 cm} (b)
 \caption{Superimposition of the identified delimiting levels on the original dendrogram (a) and the hierarchical clusters (b) detected in Figure~\ref{fig:case_den1}.}\label{fig:case_den2}
\end{figure}

First, we illustrate how to estimate the delimiting levels and hierarchical clustering organization of a given dendrogram. Figure~\ref{fig:case_den1} presents, for increasing values of the scale variable $s$, the identified delimiting levels and associated clusters, respectively to a dendrogram with 30 leaves. In order to visualize possible groups related to the considered dendrogram, a weight matrix has been associated to the data elements in the dendrogram by considering the \emph{cophenetic distance} (e.g.~\cite{sokal1962}) at each of the merging levels. Observe that this visualization is just an approximation of a possible original data set used here for illustrative purposes, not being part of the clustering methodology itself (which depends only on the associated dendrogram). The positions of the data elements in this figure and also in Figure~\ref{fig:case_den2} have been visualized using the Fruchterman-Reingold methodology~\cite{fruchterman1991}, considering the possible weight matrix obtained as described above.

As can be seen, three successive delimiting levels have been obtained, leading to the identification of the respective groups at increasing scales of $s$. Each subsequent level of clustering defines new subgroups corresponding to mergings between previously detected branches, associated to a larger scale partitioning of the original data elements.

Figure~\ref{fig:case_den2} presents the three delimiting levels in the previous examples superimposed on the original dendrogram (a), as well as the hierarchical clusters obtained at the three respective scales of detail (b). The estimated overall modularity $M$ of the dendrogram is also indicated in (b).

\begin{figure}[h]
  \centering
     \includegraphics[width=0.6 \textwidth]{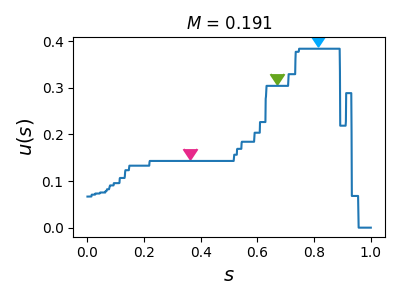}
 \caption{The signature of infinitesimal modularity obtained for the dendrogram in Figure~\ref{fig:case_den2}.}\label{fig:u_modularityxS}
\end{figure}

The infinitesimal modularity signature obtained for the dendrogram considered above is presented in Figure~\ref{fig:u_modularityxS} which indicates that, in this case, the modularity tends to increase with $s$. In addition, the plateaus observed along $s$ tend to correspond to the detected groups.

\begin{figure}[h]
  \centering
     \includegraphics[width=0.6 \textwidth]{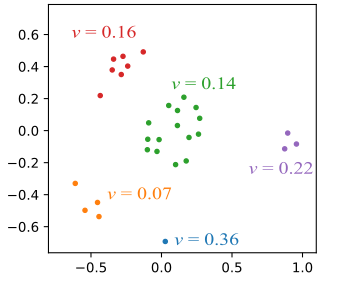}
 \caption{The individual modularity of each of the groups identified in Figure~\ref{fig:case_den1} (f).}\label{fig:individual_modularity}
\end{figure}

Figure~\ref{fig:individual_modularity} illustrates the individual modularity of each of the four identified clusters in Figure~\ref{fig:case_den1} (f).

It should be observed that the obtained individual modularity values reflect three main aspects, namely: (i) the balance between the number of elements of the paired clusters, including the dispersion of elements within the paired cluster; (ii) the effective number of elements in the involved clusters; and (iii) the distance between merged of clusters considered along the whole respective interval of the scale variable. For instance, the red and green clusters in Figure~\ref{fig:individual_modularity} present moderate individual modularity because they are large, reasonably balanced (7 and 15 elements), and relatively distant from each other. The cluster shown in orange presents the smallest modularity not only as it is relatively small, but mainly because it is largely disperse and unbalanced when compared to the neighboring clusters. The cluster shown in purple presents a relatively high individual modularity mainly as a consequence of its distance to the other clusters. The cluster in blue, which contains a single data element, presents the largest modularity because it is compact and is located at a substantial distance to the other clusters. Observe that the fact of being a single element implies null dispersion inside this group.

\begin{figure}[h]
  \centering
     \includegraphics[width=0.9 \textwidth]{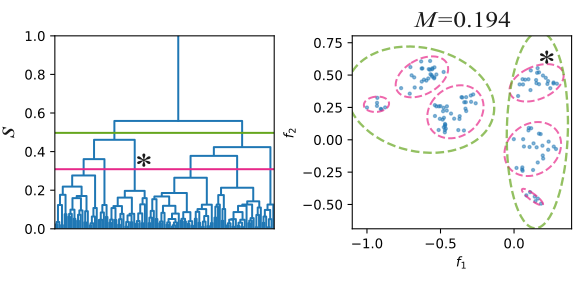}\\
     \hspace{.8 cm} (a) \hspace{5.2 cm} (b)
 \caption{Example of a dendrogram (a) obtained from the distribution of data elements shown in (b) by using the average linkage criterion, presenting the two identified delimiting levels (a) and respective hierarchical clusters (b). The sub-dendrogram indicated by the asterisk has been further studied as shown in Figure~\ref{fig:case_data1_sub}.}\label{fig:case_data1}
\end{figure}

In the following, a case-example is presented which starts from a given dataset, identified by the blue points in Figure~\ref{fig:case_data1}(b), with 123 elements characterized by two features (the spatial coordinates). The respective dendrogram, presented in Figure~\ref{fig:case_data1}(a), has been obtained from the data set by using the average linkage criterion. The two delimiting levels resulting by the proposed methodology are shown in (a), superimposed on the dendrogram. The obtained hierarchical clusters and respective modularity $M$ are shown in (b). As it can be seen from the figure, the clusters have been properly identified.

\begin{figure}[h]
  \centering
     \includegraphics[width=0.9 \textwidth]{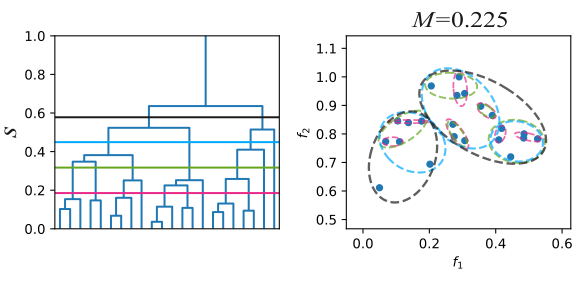}\\
     \hspace{.8 cm} (a) \hspace{5.2 cm} (b)
 \caption{Further analysis of the sub-dendrogram marked by the asterisk in Figure~\ref{fig:case_data1}(a), considering $\tilde{\sigma}=0.04$. Observe that the scale variable has been re-normalized within the interval $[0,1]$.}\label{fig:case_data1_sub}
\end{figure}

The results presented in Figure~\ref{fig:case_data1} are also considered to illustrate how the adoption of a fixed resolution $\tilde{\sigma}$ allows attention to be focused on the more separated clustering (in this particular case corresponding to those in higher hierarchy) while merging the less separated clusters. If required, each of the detected sub-dendrograms can be further studied recursively at higher resolutions, by applying the reported methodology separately to each of them. This possibility is illustrated in Figure~\ref{fig:case_data1_sub} respectively to the sub-dendrogram marked with an asterisk in Figure~\ref{fig:case_data1}(a). In this case, the scale variable for this sub-dendrogram has been re-normalized within the interval $[0,1]$ by linearly transforming the value of $s$ where the delimiting level cuts the dendrogram into $1.0$.

The recursive approach mentioned above may consider a fixed value of the smoothing parameter $\tilde{\sigma}$, or progressively smaller respective values aimed at enhancing the resolution at each recursion.

\begin{figure}[h]
  \centering
     \includegraphics[width=0.9 \textwidth]{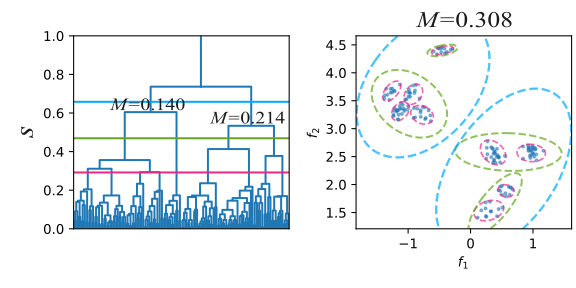}\\
     \hspace{.8 cm} (a) \hspace{5.2 cm} (b)
 \caption{Example of a dendrogram (a) obtained by the average linkage criterion from the distribution of data elements shown in (b), presenting the three identified delimiting levels (a) and respectively obtained hierarchical clusters (b). }\label{fig:case_data2}
\end{figure}

To conclude this section, another example is presented that also involves a dendrogram obtained from a respective set of initial elements shown as blue dots in Figure~\ref{fig:case_data2}(b). Now, as shown in (a), three delimiting levels have been obtained from which the hierarchical clusters in (b) have been properly identified. The modularity $M$ obtained is also indicated in (b).

\section{DLA Dendrograms}

The examples presented above concern ultrametrics dendrograms based on distances or similarities between pairs of data elements. An example of the application of the concepts and methods described in this work to a situation involving additive dendrogams of metric nature is provided in the present section, considering dendrograms obtained from two-dimensional DLA structures.

\begin{figure}[h!]
  \centering
     \includegraphics[width=0.7 \textwidth]{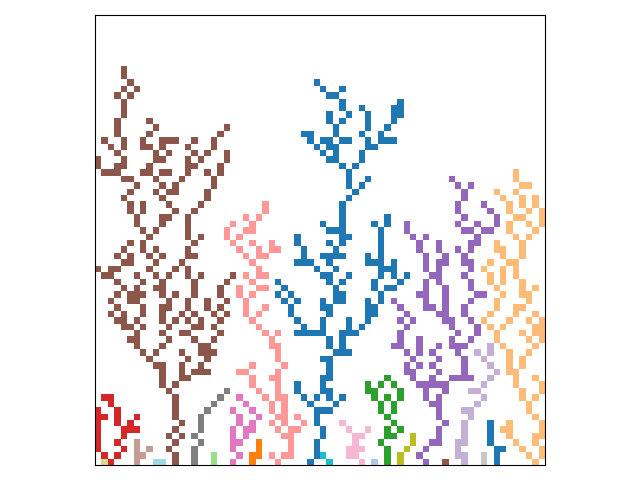}\\
     (a)\\ \vspace{0.2cm}
     \includegraphics[width=0.32 \textwidth]{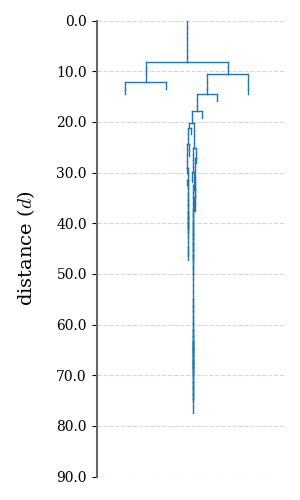}
     \includegraphics[width=0.32 \textwidth]{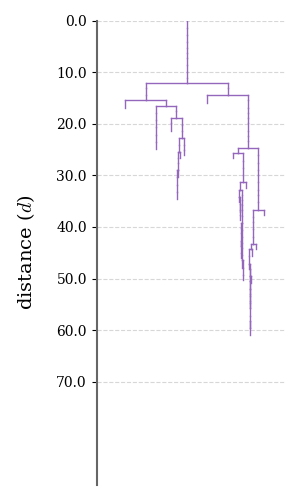}
     \includegraphics[width=0.32 \textwidth]{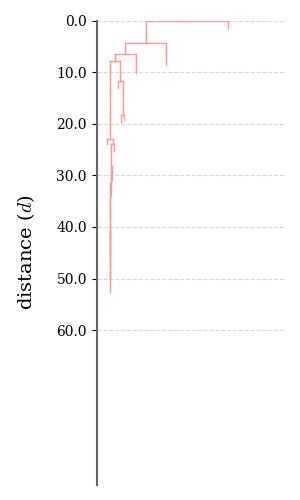}\\
     (b)
 \caption{DLA structure (a) obtained in a $70 \times 70$ lattice after 900 particles have been aggregated. Additive dendrograms obtained for the three groups marked in respective colors (b).}\label{fig:dla}
\end{figure}

Diffusion limited aggregation, or simply DLA (e.g.~\cite{meakin1995DLA}), is a theoretical model of pattern formation in which particles are progressively, after a random walk, incorporated on a growing structure. In typical 2D configurations, the growth takes place at a rectangular lattice of dimension $A \times B$, resulting in tree patterns whose roots are anchored at the bottom of the lattice. Figure~\ref{fig:dla} illustrates this type of DLA structure obtained for a $70 \times 70$ lattice.

Interestingly, the type of structure in Figure~\ref{fig:dla}(a) can be separated into several dendrograms. In the present work, this has been achieved as follows. First, each new seeded root is assigned a respective label, and each incorporated particle is labeled with the same value as the structure which has been touched. In case a particle touches simultaneously two or more already aggregated particles, one of the respective labels is drawn with uniform probability. At the end of the growth procedure, a set of connected structures can be separated in terms of the respective labels.

Each of the obtained components can be understood as an additive dendrogram, taking into account the arc-lengths between successive merging points or branches. These dendrograms can be extracted by following their elements in a sequential manner, starting from each root, unique to each component. The elements at the next height are identified, and verified for adjacency with the underneath elements. The adjacent elements are incorporated into the growing tree structure. In the case of an element being adjacent to more than a single element underneath, one of them is selected in randomly uniform manner.

Figure~\ref{fig:dla}(b) illustrates three of the dendrograms obtained from the DLA shown in (a). These structures are characterized by a few very short branches and some longer branches. At least in the case of these examples, one branch tends to dominate (in length) the overall dendrogram, corresponding to long strings of elements along the obtained DLA groups.

\begin{figure}[h]
  \centering
     \includegraphics[width=0.6 \textwidth]{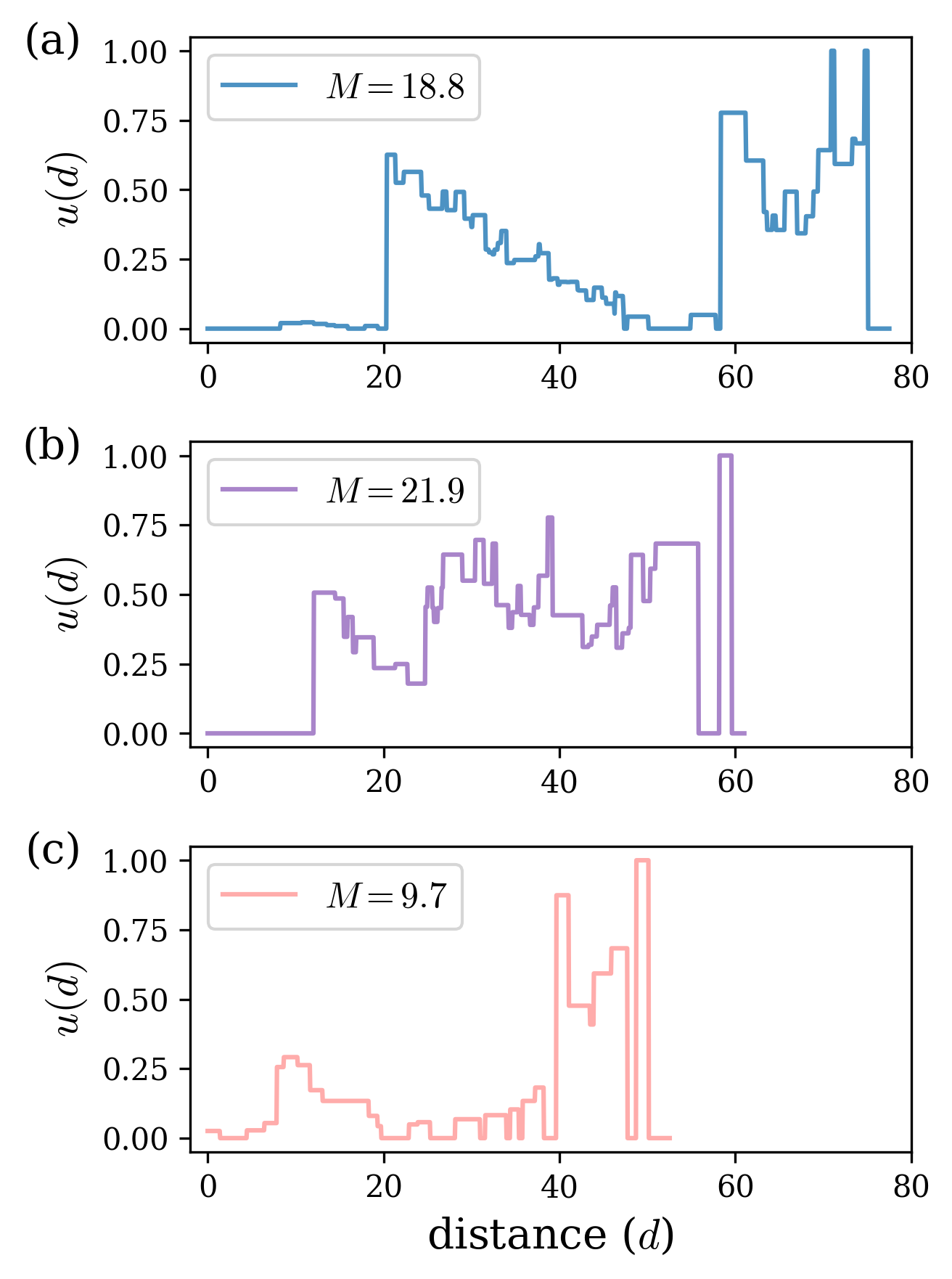}
 \caption{The infinitesimal modularity signatures $u(d)$ obtained for each of the DLA dendrograms in Figure~\ref{fig:dla}(b). Also shown are the values of three respective overall modularity $M$.}\label{fig:density_ud}
\end{figure}

The modularity of the three considered dendrograms, obtained by using the concepts and methods described in the present work, have been estimated by using infinitesimal modularity obtained for the original absolute scale of distances associated to the lengths of the segments in the structures. The infinitesimal modularity signatures in terms of distance, $u(d)$, obtained for each of the three dendrograms in Figure~\ref{fig:dla}(b), as well as their overall modularity values, are presented in Figure~\ref{fig:density_ud}.

The obtained infinitesimal modularity signatures provide information about the progression of the modularity along the distance scale $d$. For instance, the signature in Figure~\ref{fig:density_ud}(a) indicates that, after initially reaching a relatively high value of modularity in an abrupt manner (appearance of the first branches), the infinitesimal modularity subsequently faded to a near zero value, followed by marked increase of modularity implied by the last branches. Similar characteristics can be observed in the two other signatures. The largest overall modularity was obtained for the DLA structure shown in violet in Figure~\ref{fig:dla}(a), while the dendrogram in (c) presents the smallest overall modularity. It is interesting to observe that the overall modularity is not related to the complexity of the respective structures (a highly regular dendrogram can have large overall modularity, while remaining simple).

\section{Concluding Remarks}

Concepts and methodologies for identifying hierarchical clusters directly from dendrograms and estimating their hierarchical modularity have been described in this work, which are based on two main elements: (a) estimation of merging density functions; and (b) their balancing so that higher hierarchy branches are considered with enhanced weights. A possible relationship between the proposed approach and Bayesian Decision Theory has also been briefly discussed.

The main interesting features of the described approaches include: (i) performed in non-supervised manner, not requiring the estimation of the number of existing clusters; (ii) detection of hierarchical clusters directly from the equalized merging density $p(s)$ and the dendrogram, without resourcing to the original measurements associated to each data element; (iii) estimation of the several types of modularity from the infinitesimal modularity defined for a specific value of the scale variable $s$. These quantities include the average, overall, and group modularity. In addition, as has been illustrated, it is also possible to apply both the described approaches recursively to sub-dendrograms of specific interest.

The influence of the parameter $\tilde{\sigma}$ on cluster identification and modularity estimation has also been briefly addressed, and preliminary results indicate that, at least for the type of data and configurations considered, relatively little influence could be observed in both cases for relatively small variations of that parameter.

The potential of the described approaches has been illustrated for some types of dendrograms and data sets, as well as to the analysis of additive dendrograms obtained from DLA structure. The results obtained indicate proper detection of clusters and modularity estimation, at least for the dendrograms and configurations considered.

To a good extent, the encouraging performance observed for the described cluster detection approach stems from the consideration of the fact that dendrograms (and respective datasets) characterized by multiple modularity scales are unlikely to have their groups effectively identified while considering a single slicing through the dendrogram (single scale). By estimating the several hierarchical levels in an unsupervised manner, the proposed approach automatically adapts the identification of the groups with respect to those levels.

In addition, it should be observed that dendrogram modularity has been here approached in terms of a single overall value, rather than estimating the modularity among specific subclusters. While both approaches are interesting, the hierarchical modularity considered in the present work does not rely on cluster identification, which is performed independently (though being influenced by the overall modularity), instead of trying to identify, at a single modularity scale.

The adopted separation between modularity maximization and cluster identification has ultimately allowed a particularly direct and computationally effective computations. Several types of hierarchical modularity have been described, based on the concept of infinitesimal modularity, including the individual, set, overall, and groups modularity. Each of these types of modularity can be used to quantify specific aspects of the hierachical-modular structure of given dendrograms.

The reported results indicate that the proposed concepts and methodologies have promising potential for performing non-supervised pattern recognition and estimating hierarchical modularity. In addition, the approach is relatively simple, involving low complexity order (the cluster identification is performed along the one-dimensional scale variable $s$). However, the results obtained are specific to the types and sizes of data sets, the number and shapes of clusters, the types of dendrograms, the linkage criterion, and the parametric configurations considered. Therefore, it would be necessary to further investigate the performance of the methodology in more general cases and configurations.

Additional perspectives for further investigations include, but are not limited to, the development of approaches for generating dendrograms by using the reported concepts; studying the effect of the fixed smoothing parameters $\tilde{\sigma}$ more systematically; and developing measurements (e.g.~functionals or multi-resolution indicators) as a means to characterize additional properties of dendrograms directly from their respective equalized merging density functions. Another possible development consists of adapting the proposed cluster detection and modularity estimation approaches to graphs and complex networks represented by their connectivity matrices. Actually, the reported approach can be potentially adapted and applied to study any type of abstract or real structure and data that can be effectively represented in terms of respective dendrograms, including physical branched structures, including hydrographic systems, and neuronal cells.

\section*{Acknowledgments}
A. Benatti is grateful to MCTI PPI-SOFTEX (TIC 13 DOU 01245.0102\\22/2022-44), FAPESP (grant 2025/26083-7 and 2022/15304-4). Luciano da F. Costa thanks CNPq (grant no.~313505/2023-3) and FAPESP (grant 2022/15304-4).

\bibliography{ref}
\bibliographystyle{unsrt}

\end{document}